\documentclass[useAMS,usenatbib]{mn2e}
\usepackage{graphics}
\usepackage{amssymb}
\usepackage{epsfig}

\title[Photometry of irregular dwarfs within 10 Mpc]{$B,V,I$ -- photometry of 
20 dwarf irregular galaxies within 10 Mpc}
\author[L.N. Makarova et al.]{L. Makarova,$^{1,2}$ \thanks{email lidia@sao.ru}
I. Karachentsev,$^{1}$ L. Rizzi,$^{3}$ R. B. Tully,$^{4}$ G. Korotkova$^{1}$\\
$^{1}$Special Astrophysical Observatory, Nizhnij Arkhyz, 
Karachaevo-Cherkessia, Russia 369167 \\
$^{2}$Isaac Newton Institute of Chile, SAO Branch \\
$^{3}$Joint Astronomy Centre, 660 N. A'ohoku Place, University Park, 
Hilo, HI 96720, USA \\
$^{4}$Institute for Astronomy, University of Hawaii, 2680 Woodlawn Drive,
Honolulu, HI 96822, USA}
\begin{document}

\maketitle

\label{firstpage}

\begin{abstract}
CCD -- photometry is presented for 20 dIrr galaxies situated in
the nearby complexes CenA/M83, and CVnI as well as in the general field
of the Local Volume. We present integrated magnitudes of the galaxies in
$B,V,I$- bands and also surface brightness profiles to a median
isophote  $\mu_B$ $\sim$ 28 mag sq. arcsec$^{-1}$. 
The popular Sersic parameterization of surface brightness profiles
generally does a poor job of simultaneously fitting the inner cores and
outer exponential surface brightness fall-offs observed in many of
our targets.  The observed sample is
a part of a general project to image about 500 nearby 
(D $<$ 10 Mpc) dwarf galaxies in multiple bands.
\end{abstract}


\section{Introduction}

Dwarf galaxies in the Local Volume ( D $\leq$ 10 Mpc) are well
recognised now as important objects for our understanding of such 
diverse topics as star formation, galaxy evolution,
and large scale structure. Many nearby galaxies are well resolved into
individual stars, which gives a possibility to estimate their
distances via the luminosity of the tip of Red Giant Branch (TRGB).
The study of the nearby  universe allows
us to probe the cosmic structure and galaxy properties at
the present epoch and thus serves as a reference for evolutionary studies.
Recent systematic surveys for dwarf galaxies have doubled the number 
of objects known in the Local Volume (=LV) which are collected in the Catalogue
of Neighbouring Galaxies (= CNG, Karachentsev et al. \cite{k04}).
{\bf Recent discoveries (see, for example, Belokurov et al. \cite{belokurov},
Koposov et al. \cite{koposov}) also increase extremely the number 
of known dwarf galaxies in our Local group.}
Unfortunately, the discovery of large numbers of nearby galaxies, mostly
of low surface brightness, has not been accompanied by systematic
photometric studies. As a result, many new members of the LV so far have
only rough estimates of their apparent magnitudes.
There is no homogeneous optical survey embracing
galaxies of both northern and southern hemispheres. The uniform
survey 2MASS was performed in the near infrared 
bands (Jarrett et al. \cite{jarrett}). However, it was
insensitive to objects of low
surface brightness because of short exposure times and high sky backgrounds. 
Most LV galaxies are not represented in 2MASS. 
{\bf About 65\% of our sample galaxies are presented in the SDSS, 
but the exposures are too short to analyze the low surface brightness objects.}
At present we are faced
with the strange situation that the majority of the LV galaxies have
radial velocities measured with $\sim$1\% accuracy and distances determined
with an error $\sim$10\%, but the integrated magnitudes of many nearby 
galaxies have uncertainties of 0.5 mag. This discordance makes it difficult
to study the faint end of the luminosity function in detail. 
The main photometric parameters, such as total magnitudes and colours,
remain poorly known for many nearby dwarf galaxies.
A study of these galaxies not only in visual bands ($B$ and $V$
Johnson-Cousins filters), but also in the near infrared
($I$ Cousins filter), where extinction is less, can make it
possible to investigate in detail the galaxy morphology
and structure of these galaxies.

This is the fourth paper in a series of articles that present
results from the observations of nearby dwarf galaxies, including presentation
of their general photometric parameters. In the previous papers (Makarova
\cite{m98}, Makarova et al. \cite{m02}, \cite{m05}) we described photometry
of 78 nearby galaxies. Here we present photometric data for 20
more dwarf galaxies observed with the University of Hawaii 2.2m telescope.

\begin{table*}
\caption{General parameters of the sample}
\medskip
\begin{tabular}{lrcccccccc}
\hline
Name  &       Leda No. &     RA  &     Dec &  D & $A_B$& $V_h$  & $V_{LG}$& Dist.& Method \\ 
      &                &         &         &arcmin&    &km s$^{-1}$& km s$^{-1}$& Mpc & \\ \hline
KKH 46     &  2807128 & 09 08 36.6 &  +05 17 32 & 0.7& 0.20&  598 & 409 & 6 & CNG, h \\
D634--03   &  2806961 & 09 08 53.5 &  +14 34 55 & 0.4& 0.16&  319 & 173 &9.5 & [1], trgb \\
Antlia     &  029194  & 10 04 04.0 & --27 19 55 & 2.0& 0.34&  362 &  66 &1.3 & [2], trgb \\
KKH 60     &  2807134 & 10 15 59.4 &  +06 48 17 & 0.8& 0.09&  290:& --  & --  & -- \\
UGC 5672   &    30818 & 10 28 20.9 &  +22 34 17 & 1.8& 0.10&  531 & 428 &6  & CNG, bs \\
KKs44      &    36014 & 11 37 53.2 & --39 13 13 & 1.4& 0.62&  654 & 362 &6.1 & [3], trgb \\
KK 144     &   166137 & 12 25 27.9 &  +28 28 57 & 1.5& 0.11&  483 & 453 &6  & CNG, h \\
KK 149     &    41093 & 12 28 52.3 &  +42 10 40 & 0.8& 0.11&  407 & 446 &6  & CNG, h \\
KK 151     &    41314 & 12 30 23.8 &  +42 54 05 & 1.2& 0.08&  436 & 479 &7  & CNG, h \\
KK 160     &   166142 & 12 43 57.4 &  +43 39 41 & 0.8& 0.11&  293 & 346 &5  & CNG, h \\
ESO 381--018&   42936 & 12 44 42.7 & --35 58 00 & 1.2& 0.27&  610 & 353 &5.3 & [3], trgb \\
ESO 381--020&   43048 & 12 46 00.4 & --33 50 17 & 3.0& 0.28&  585 & 332 &5.4 & [3], trgb \\
KKSG 37    &  3097714 & 12 48 01.0 & --12 39 22 & 0.6& 0.22&   84:&  -- & --  & -- \\
ESO 443--09&    43978 & 12 54 53.6 & --28 20 27 & 0.8& 0.28&  645 & 410 &5.5 & [3], trgb \\
DDO 155    &    44491 & 12 58 40.4 &  +14 13 03 & 1.1& 0.11&  214 & 136 &2.1 & CNG, trgb \\
PGC 45628  &    45628 & 13 09 36.6 & --27 08 26 & 0.6& 0.33&  693 & 470 &6  & CNG, h \\
PGC 170257 &   170257 & 13 29 21.0 & --21 10 45 & 0.6& 0.46&  650 & 457 &6  & CNG, h \\
UGC 8651   &    48332 & 13 39 53.8 &  +40 44 21 & 2.3& 0.03&  202 & 272 &3.0 & [2], trgb \\
UGC 8760   &    49158 & 13 50 51.1 &  +38 01 16 & 2.2& 0.07&  191 & 257 &3.2 & [2], trgb \\
UGC 8833   &    49452 & 13 54 48.7 &  +35 50 15 & 0.9& 0.05&  226 & 285 &3.2 & [2], trgb \\
\hline
\multicolumn{10}{l}{[1] Karachentsev et al. 2006}\\
\multicolumn{10}{l}{[2] Tully et al. 2006}\\
\multicolumn{10}{l}{[3] Karachentsev et al. 2007}\\
\end{tabular}
\end{table*}

\section{Observations}
The galaxies were selected from the CNG in accordance with the observation
season. About 2/3 of the galaxies turn out to be members of two nearby
complexes: CenA/M83 group and Canes Venatici I cloud, the remaining ones
lie in the general field. Some basic
parameters of the sample galaxies are presented in Table~1.
The columns are:

\noindent
{\it Column 1:} galaxy name.

\noindent
{\it Column 2:} LEDA number.

\noindent
{\it Column 3and 4:} Right ascension and Declination (2000).

\noindent
{\it Column 5:} Major angular diameter from the CNG in arcminutes.

\noindent
{\it Column 6:} galactic absorption in $B$ band taken from
IRAS/DIRBE maps (Schlegel, Finkbeiner, Davis \cite{schlegel}). Galactic
absorption in $V$ and $I$ can be computed as follow: $A_V = 0.768 A_B$
and $A_I = 0.449 A_B$.

\noindent
{\it Column 7:} heliocentric radial velocity in km s$^{-1}$.

\noindent
{\it Column 8:} The Local Group centroid radial velocity

\noindent
{\it Column 9:} galaxy distance in Mpc.

\noindent
{\it Column 10:} source of the distance and method of its estimation: h -- 
radial velocity; trgb -- tip of the red giant branch; bs -- brightest 
supergiants.

Our targets were observed using the 2.2m
University of Hawaii telescope at Mauna Kea Observatory,
Hawaii, USA on February 17--22, 2004. We carried out direct
imaging of the galaxies using the Optic CCD Camera, which gives a
9.3'$\times$9.3' field of view with the pixel scale of 0.14''/pix.
The galaxy images were obtained using $B$, $V$ and $I$ Johnson-Cousins filters
with typical exposure time of 1200 sec, 900 sec and 600 sec,
respectively. Bias frames, twilight flats and standard fields
from Landolt \cite{landolt} were observed each night for the calibration. 

\section{Data reduction}

\subsection{Initial processing}
The images were processed using the {\it ccdproc} package within IRAF. 
Initial processing of the data contained usual steps such
as bias and dark frame subtraction and flat-fielding.
Further reductions were made with the MIDAS package developed by ESO.
Cosmic ray hits were removed with the FILTER/COSMIC procedure.
The galaxy images obtained in the same
filter were then co-added. The resulting $B$-band images of the galaxies
under study are presented in Fig.~1. The picture size is 3.5'$\times$3.5', 
North is up and East is left.

\begin{figure*}
\psfig{figure=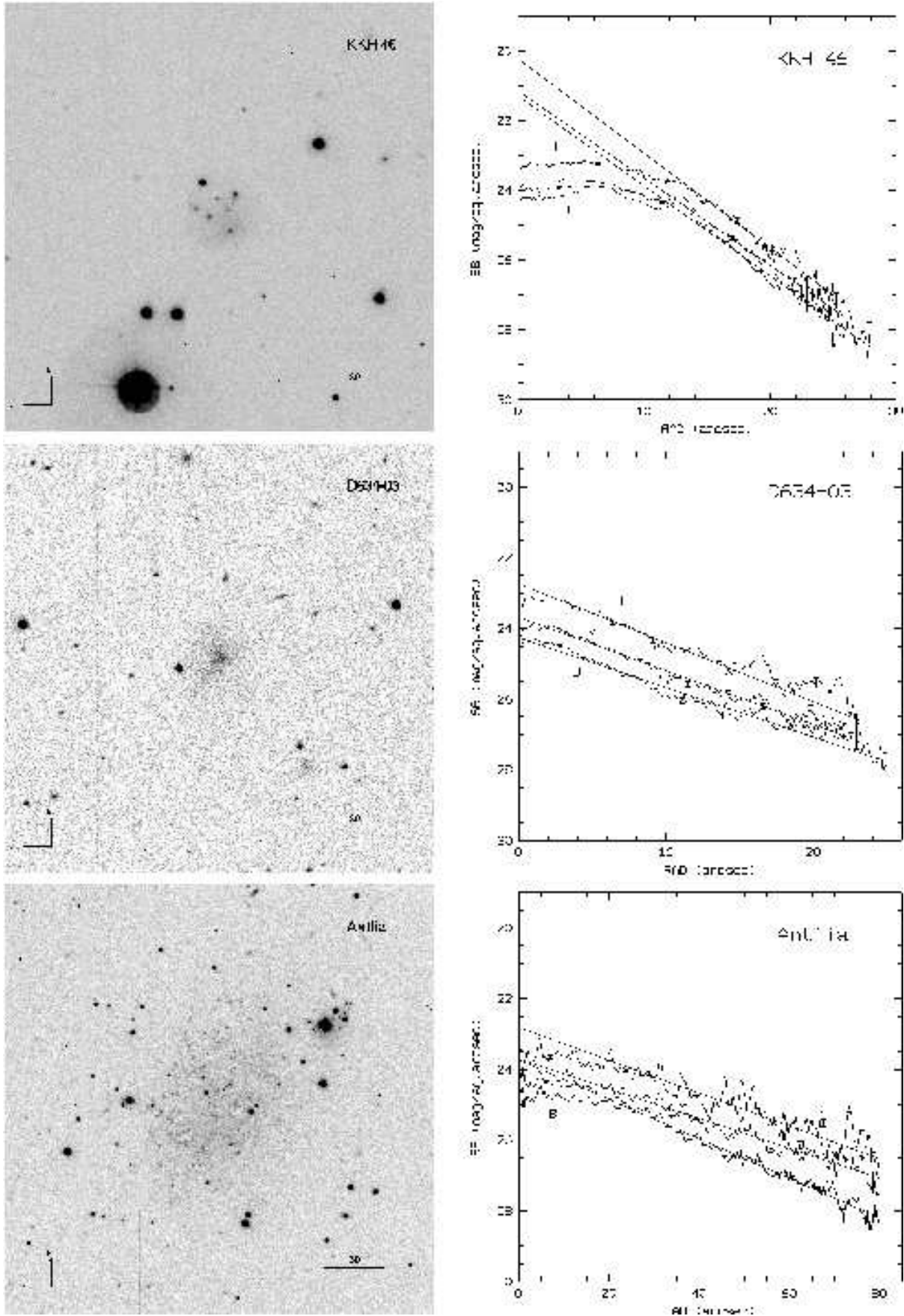,width=16cm}
\caption{B-band images and B, V, I surface brightness profiles 
of the galaxies under study. The picture sizes are
3.5'$\times$3.5', North is up and East is left.}
\end{figure*}

\setcounter{figure}{0}
\begin{figure*}
\psfig{figure=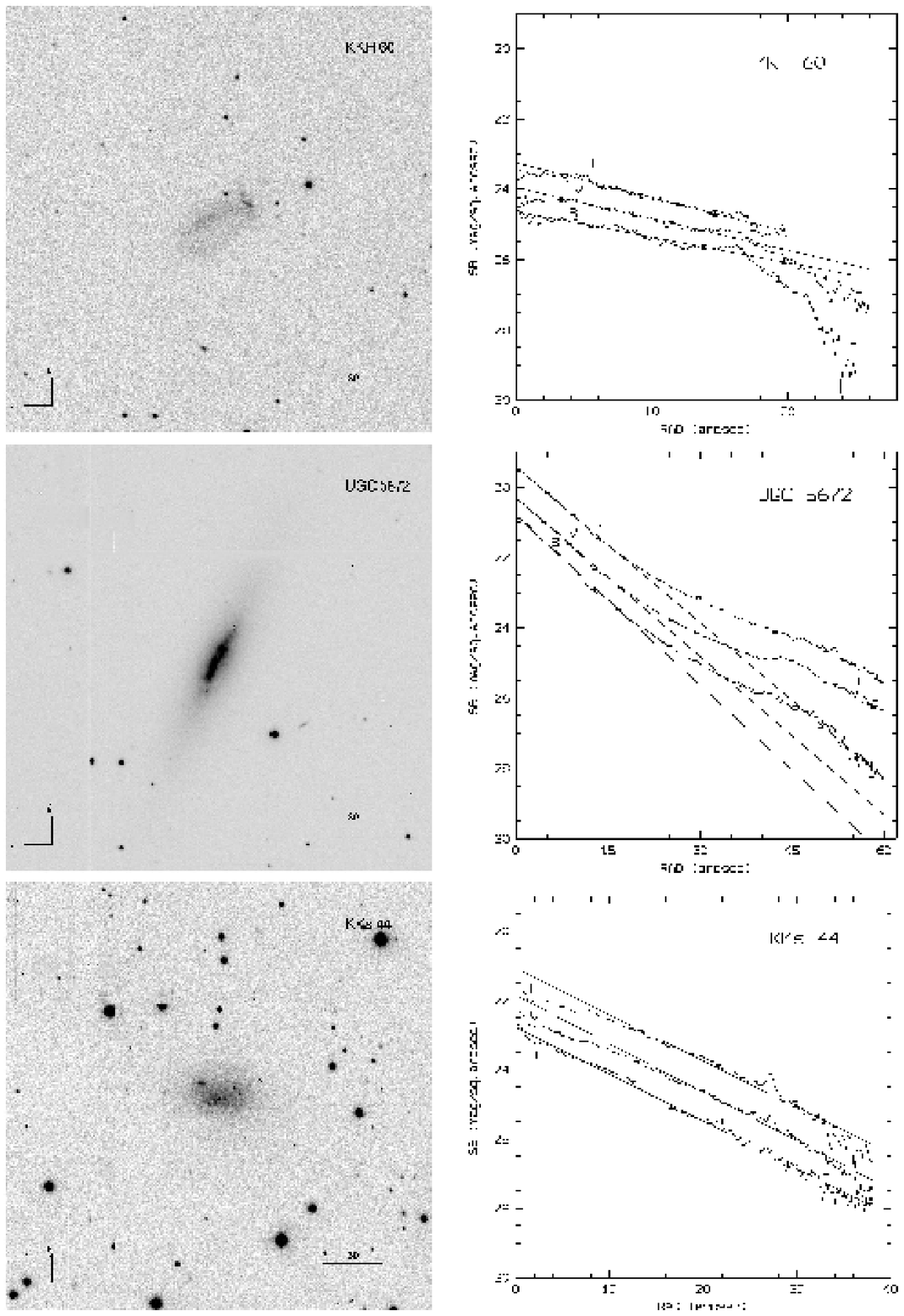,width=16cm}
\caption{continued.}
\end{figure*}

\setcounter{figure}{0}
\begin{figure*}
\psfig{figure=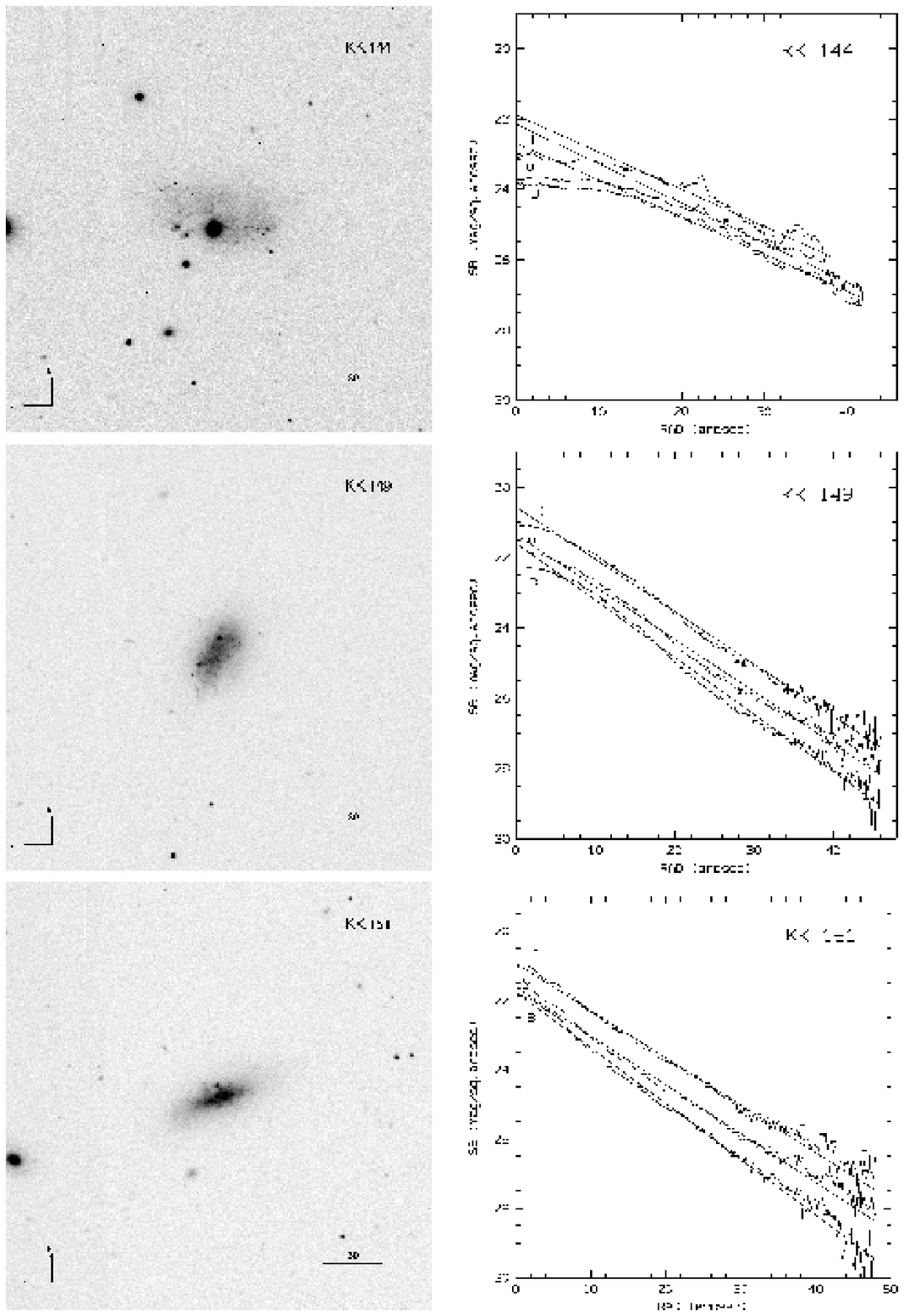,width=16cm}
\caption{continued.}
\end{figure*}

\setcounter{figure}{0}
\begin{figure*}
\psfig{figure=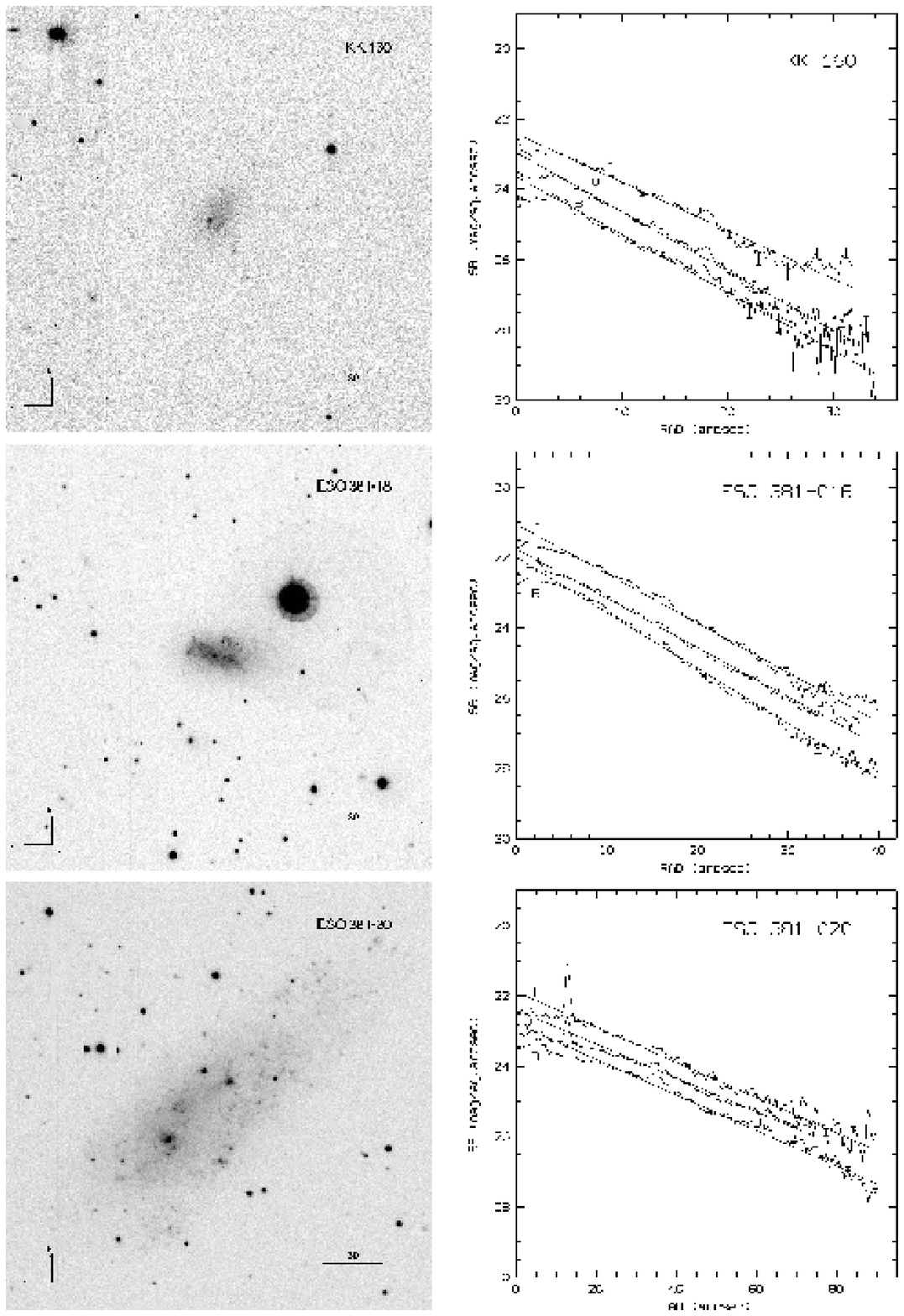,width=16cm}
\caption{continued.}
\end{figure*}

\setcounter{figure}{0}
\begin{figure*}
\psfig{figure=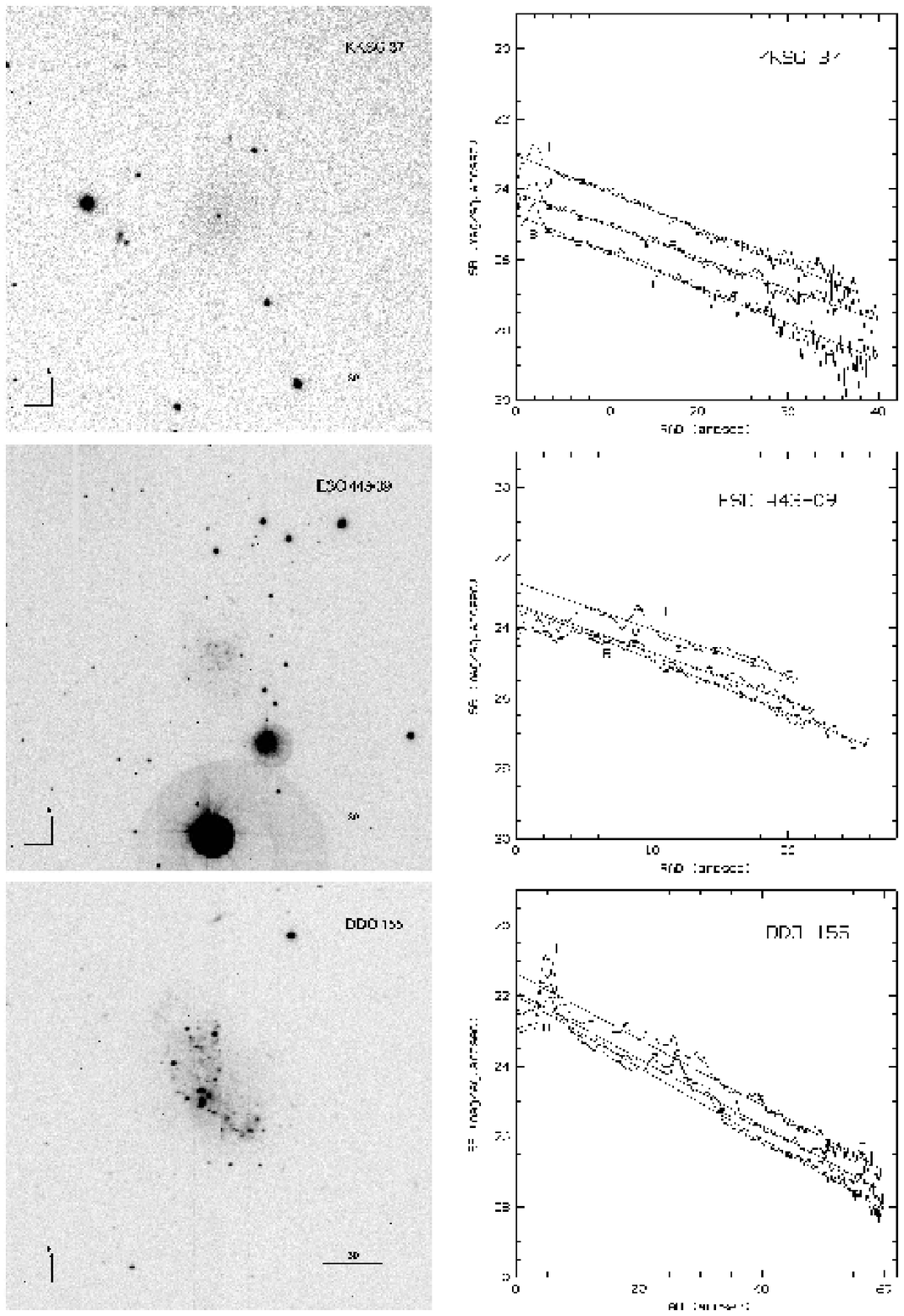,width=16cm}
\caption{continued.}
\end{figure*}

\setcounter{figure}{0}
\begin{figure*}
\psfig{figure=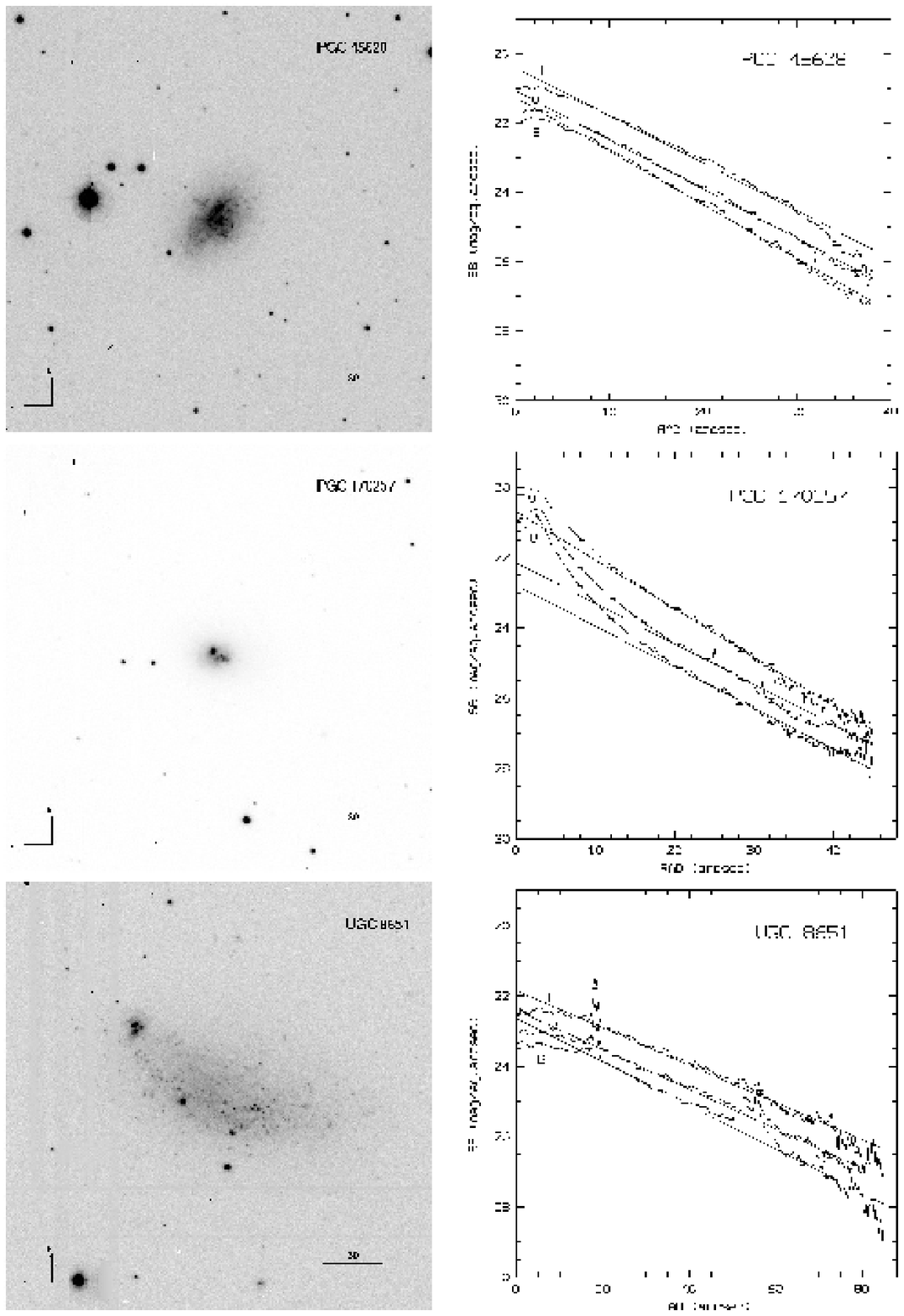,width=16cm}
\caption{continued.}
\end{figure*}

\setcounter{figure}{0}
\begin{figure*}
\psfig{figure=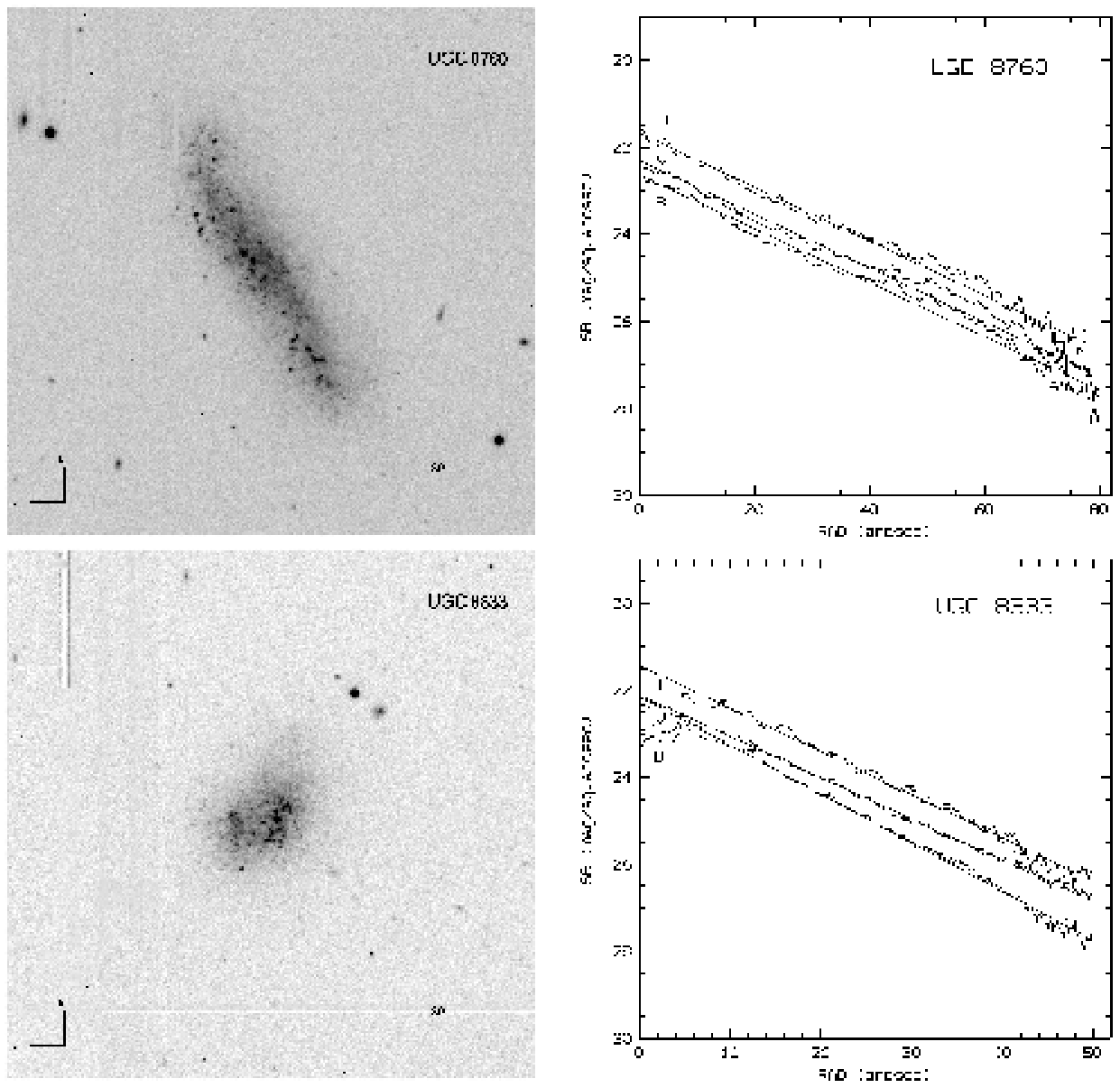,width=16cm}
\caption{continued.}
\end{figure*}

\subsection{Sky background determination}
Background stars were removed from the frames by fitting a second-degree 
surface in circular pixel-area. The sky background was then 
approximated by a tilted plane, created from a two-dimension polynomial,
using the least-squares method (FIT/FLAT\_SKY). The accuracy of the sky
background determination is about 0.6 -- 1.0 \% of the original sky
level. A typical value of the sky surface brightness is near 23 
mag sq.sec$^{-1}$ in $B$ band. Therefore, the mean error of sky 
determination does not exceed 0.14 -- 0.23 mag.

\subsection{Photometric calibration}
To transform the
instrumental magnitudes to the standard Johnson-Cousins system,
zero points and colour coefficients were determined from the standard
stars observed in the clear photometric nights at February 17, 18 and
20. We used the mean Mauna Kea summit extinction
coefficients for the transformation due to nonstable result 
of these coefficient measurements
from our data. Too few stars were observed to constrain
such measurements. The observational nights at February 21 and 22 
were
partially non photometric. Therefore, we used
zero points obtained from the appropriate standard field closest 
to the time of the respective observations. These zero points have
good agreement, the additional uncertainty in the calibration will not 
exceed 0.03 mag in $B$, and 0.02 mag in $V$ and $I$.
The uncertainty of the overall transformation 
to the standard system is 0.01 mag in the $B$, $V$ and $I$ filters.

\begin{table*}
\caption{Total magnitudes of the dwarf galaxies}
\medskip
\begin{tabular}{lccccccccr}
\hline
Name    &        $B_T$ &  $V_T$ &  $I_T$&  $B-V$& $V-I$ & $D_{25}$ & $D_{Holm}$ & $B_{Holm}$ & $M_B$ \\
        &        mag   &  mag   &  mag  &   mag &  mag  & arcsec   & arcsec     &  mag       & mag   \\ \hline
KKH 46       &    16.96&  16.72 & 16.27 &  0.24&   0.45&  28.8& 41.2 &16.97 & --12.02 \\
D634--03     &    18.27&  17.78 & 16.92 &  0.49&   0.86&  11.0& 28.8 &18.45 & --11.77 \\
Antlia       &    15.57&  15.16 & 14.60 &  0.41&   0.56&  47.6& 92.6 &15.70 & --10.37 \\
KKH 60       &    17.76&  17.24 & 16.64 &  0.52&   0.60&  12.0& 28.4 &17.90 & --    \\
UGC 5672     &    14.48&  13.69 & 12.79 &  0.79&   0.90&  60.0& 98.0 &14.52 & --14.62 \\
KKs44        &    16.30&  15.74 & 14.71 &  0.56&   1.03&  32.6& 57.0 &16.34 & --13.24 \\
KK 144       &    15.86&  15.53 & 14.96 &  0.33&   0.57&  46.0& 72.4 &15.91 & --13.25 \\
KK 149       &    15.54&  15.04 & 14.31 &  0.50&   0.73&  40.0& 55.0 &15.59 & --13.53 \\
KK 151       &    15.63&  15.16 & 14.49 &  0.47&   0.67&  38.8& 60.0 &15.71 & --13.58 \\
KK 160       &    17.65&  17.02 & 16.10 &  0.63&   0.92&  14.8& 36.2 & 17.87& --10.87 \\
ESO 381--018 &    15.73&  15.30 & 14.67 &  0.43&   0.63&  39.0& 68.4 & 15.73& --13.17 \\
ESO 381--20  &    14.24&  13.93 & 13.32 &  0.31&   0.61&  90.0& 150.0 &14.28& --14.72 \\
KKSG 37      &    18.58&  17.78 & 16.77 &  0.80&   1.01&   6.0& 30.0 &18.58 &  --     \\
ESO 443--09  &    17.38&  17.17 & 16.52 &  0.21&   0.65&  20.4& 39.4 &17.45 & --11.60 \\
DDO 155      &    14.79&  14.50 & 14.02 &  0.29&   0.48&  64.0& 92.4 &14.84 & --11.93 \\
PGC 45628    &    15.11&  14.71 & 14.05 &  0.40&   0.66&  47.6& 62.4 &15.12 & --14.28 \\
PGC 170257   &    15.32&  14.74 & 14.04 &  0.58&   0.70&  37.6& 60.0 &15.34 & --14.14 \\
UGC 8651     &    14.22&  13.95 & 13.34 &  0.27&   0.61&  88.0& 150.0 &14.26 & --13.20 \\
UGC 8760     &    14.47&  14.23 & 13.67 &  0.24&   0.56&  84.0& 130.0 &14.51 & --13.11 \\
UGC 8833     &    15.30&  14.97 & 14.38 &  0.33&   0.59&  51.2& 77.6 &15.38 & --12.27 \\ \hline
\end{tabular}
\end{table*}

\section{Photometric measurements}
\subsection{Total magnitudes}
To measure total galaxy magnitudes in each band, integrated
photometry was performed in circular apertures with increasing radii
from a
pre-chosen centre to the faint outskirts of the galaxies.
{\bf The centers were determined interactively,
because a significant part of the sample consists of the galaxies of irregular
morphology.}
The total magnitude was then estimated as the asymptotic value of
the obtained radial growth curve. 
The uncertainties of the total magnitude determination were 
0.05 mag in $B$, 0.05 mag in $V$ and 0.08 mag in $I$. The measurement results 
are summarised in Table~2, where the columns are:

\noindent
{\it Column 1:} galaxy name.

\noindent
{\it Column 2, 3 and 4:} total magnitudes in $B$, $V$ and $I$ band, 
respectively (not corrected for galactic absorption).

\noindent
{\it Column 5 and 6:} total colour $B-V$ and $V-I$, respectively
(not corrected for galactic absorption).

\noindent
{\it Column 7:} the standard diameter (arcsec) at the surface brightness
level of 25 mag sq. sec$^{-1}$ in $B$ band.

\noindent
{\it Column 8:} the Holmberg diameter (arcsec) corresponding to the blue
surface brightness of 26.5 mag sq. sec$^{-1}$.

\noindent
{\it Column 9:} $B$ magnitude within the Holmberg diameter

\noindent
{\it Column 10:} absolute $B$ total magnitude corrected for Galactic absorption.

{\bf Slight changes of a slope in about 75\% of the surface brightness profiles
can indicate redder colour index of the galaxies in their outer parts.
This gradient occurs probably due to increase in the average age of the
stellar population towards the edge of the galaxy.} 

\begin{table*}
\caption{The fitting parameters of the dwarf galaxies}
\medskip
\begin{tabular}{lccrccrccr}
\hline
 name & $\mu_{B}^p$ & $\mu_{B}^f$&$h_B$&$\mu_{V}^p$&$\mu_{V}^f$&$h_V$&$\mu_{I}^p$&$\mu_{I}^f$&$h_I$ \\
    & mag sq.sec$^{-1}$ & mag sq.sec$^{-1}$ & arcsec& mag sq.sec$^{-1}$ & mag sq.sec$^{-1}$ & arcsec& mag sq.sec$^{-1}$ & mag sq.sec$^{-1}$ & arcsec \\
 \hline
KKH 46      &   24.2& 21.30&  4.2& 23.8& 21.15&  4.3& 23.3& 20.23&  4.0 \\
D634-03     &   24.2& 24.27&  7.8& 23.9& 23.68&  7.1& 23.2& 22.74&  6.5 \\  
Antlia      &   24.8& 23.73& 19.4& 24.3& 23.71& 25.8& 23.7& 22.82& 23.4  \\
KKH 60      &   24.7& 24.64& 14.9& 24.4& 23.96& 12.2& 23.6& 23.25& 11.1  \\
UGC 5672    &   20.8& 20.81&  6.8& 20.3& 20.31&  7.2& 19.5& 19.44&  7.4  \\
KKs 44      &   22.6& 22.77&  8.0& 22.3& 21.86&  7.7& 21.7& 21.08&  8.1  \\
KK 144      &   23.9& 22.69& 10.3& 23.7& 22.15&  9.6& 23.1& 21.89& 10.3 \\                          
KK 149      &   22.3& 21.60&  6.7& 21.8& 21.35&  7.3& 21.1& 20.56&  7.3  \\                          
KK 151      &   22.0& 21.76&  6.7& 21.7& 21.62&  7.7& 21.1& 20.93&  7.9  \\       
KK 160      &   24.3& 23.65&  6.6& 23.6& 22.96&  6.4& 22.6& 22.42&  7.9  \\                         
ESO 381-18  &   22.6& 21.99&  6.9& 22.3& 21.76&  7.7& 21.6& 21.05&  7.7  \\                       
ESO 381-20  &   23.1& 22.79& 21.4& 23.0& 22.41& 22.5& 22.5& 21.88& 21.6  \\                              
KKSG 37     &   24.7& 24.74& 10.6& 24.2& 24.12& 12.0& 23.4& 23.04& 10.5  \\                         
ESO 443-09  &   24.0& 23.36&  7.0& 24.1& 23.33&  7.9& 23.5& 22.71&  8.2  \\                        
DDO 155     &   22.9& 21.99& 10.7& 22.6& 21.89& 11.4& 22.3& 21.36& 11.6  \\                    
PGC 45628   &   21.9& 21.26&  7.0& 21.6& 21.07&  7.6& 21.0& 20.41&  7.9  \\                         
PGC 170257  &   20.9& 22.79&  9.3& 20.6& 22.15&  9.4& 20.2& 20.69&  7.8  \\
UGC 8651    &   23.3& 22.63& 17.5& 23.1& 22.40& 19.3& 22.4& 21.85& 20.5  \\
UGC 8760    &   22.8& 22.64& 17.5& 22.4& 22.31& 17.6& 21.8& 21.68& 17.5  \\
UGC 8833    &   23.0& 22.15&  9.7& 22.8& 22.12& 11.5& 22.2& 21.43& 11.0  \\ \hline
\end{tabular}
\end{table*}

\subsection{Surface brightness profiles}

Azimuthally averaged surface
brightness profiles have been widely used (Karachentseva et al.
 \cite{karachentseva96}, Papaderos et al. \cite{papaderos},
Bremnes, Binggeli, Prugniel \cite{bremnes98}, 
Gil de Paz \& Madore \cite{gildepaz})
for the investigation of dwarf galaxies.
They allow one to improve the accuracy of surface photometry in galaxies
of low surface brightness and irregular structure.
Azimuthally averaged surface brightness profiles for our galaxies
were obtained by differentiating the galaxy growth curves with
respect to radius (Bremnes et al.\ \cite{bremnes98}). The resulting 
profiles
in the $B$, $V$ and $I$ colours are displayed in Fig.~1. Most of the galaxies
were measured up to the level of 28 mag arcsec$^{-2}$ in the $B$ filter.
Mean uncertainties of the
measurements were estimated by intercomparison of individual
profiles for the same objects obtained from different frames
in the same passband.  They amount to about 0.08 mag at the 23 mag
arcsec$^{-2}$ isophotal level and about 0.3 mag to 0.4 mag at the 27 mag
arcsec$^{-2}$ isophotal level in each of the filters.

It is well known that surface brightness profiles of dwarf galaxies
(both irregular and spheroidal) and also disks of spiral galaxies can
be fitted by an exponential intensity law of brightness distribution
( de Vaucouleurs \cite{vau})
$$ I(r) = I_0*exp(-r/h)$$
or, in magnitudes per sq.sec
$$ \mu(r) = \mu_0+1.086*(r/h), $$
where $\mu_0$ is the central surface brightness and h is the
exponential scale length.
Most of the measured galaxies are well-fitted by an exponential law
in the whole profile or in its outer parts. 
As can be seen from Fig.~1, most of the surface brightness profiles
have a linear outer part, which can be considered as evidence for
the presence of an underlying stellar disk component of the galaxy. About
70\% of the surface brightness profiles show central light depression,
which is very common for dwarf irregular galaxies. We also have five
profiles in our sample that seem purely exponential, and two of 
the galaxies have prominent light excesses in their central parts. 
Our previous sample of nearby dwarfs (Makarova et al. \cite{m05}) show similar
tendencies.

The {\bf unweighted} exponential fits to the surface brightness 
profiles were done by
linear regression. The rms deviation of the derived model profiles
from the original ones does not exceed 0.10 mag. The results of
the fitting are presented in Table~3. The columns are:

\noindent
{\it Column 1:} galaxy name.

\noindent
{\it Column 2:} central surface brightness in $B$ filter estimated from
the original profile.

\noindent
{\it Column 3 and 4:} best-fitting parameters of the exponential disk in $B$
filter (the central surface brightness $\mu_0$ (mag sq.sec$^{-1}$)
and exponential scale length h (arcsec)).

\noindent
{\it Column 5:} central surface brightness in $V$ filter estimated from
the original profile.

\noindent
{\it Column 6 and 7:} best-fitting parameters of the exponential disk in $V$
filter.

\noindent
{\it Column 8:} central surface brightness in $I$ filter estimated from
the original profile.

\noindent
{\it Column 9 and 10:} best-fitting parameters of the exponential disk in $I$
filter.

The listed surface brightnesses were not corrected for the Galactic absorption.

\subsection{Photometric uncertainties} 
Summing up
all internal errors given above, we obtain the resulting errors
of the total magnitude estimation to be about 0.24$^m$ in $B$, $V$ and $I$ band, 
and the resulting errors of surface brightness
estimation to be about 0.24$^m$ at the 23 mag sq.sec$^{-1}$ 
isophotal level and
about 0.38-0.46 at the 27 mag sq.sec$^{-1}$ isophotal level 
in each photometric band.

External photometric errors can be estimated by a comparison of
the total magnitudes of the observed galaxies with the 
published magnitudes of
these galaxies.
The comparison of the total $B$-magnitudes estimated in the present work
with those given in the articles of Bremnes, Binggeli, Prugniel \cite{bremnes99},
Hopp\&Schulte-Ladbeck \cite{hs95}, Makarova et al. \cite{metal98} and 
Makarova et al. \cite{m02} for five galaxies
in common, yields the satisfactory agreement of -0.02$\pm$0.15 mag.
We did not consider other bands due to poor statistic (one-two 
published values).

\begin{figure*}
\psfig{figure=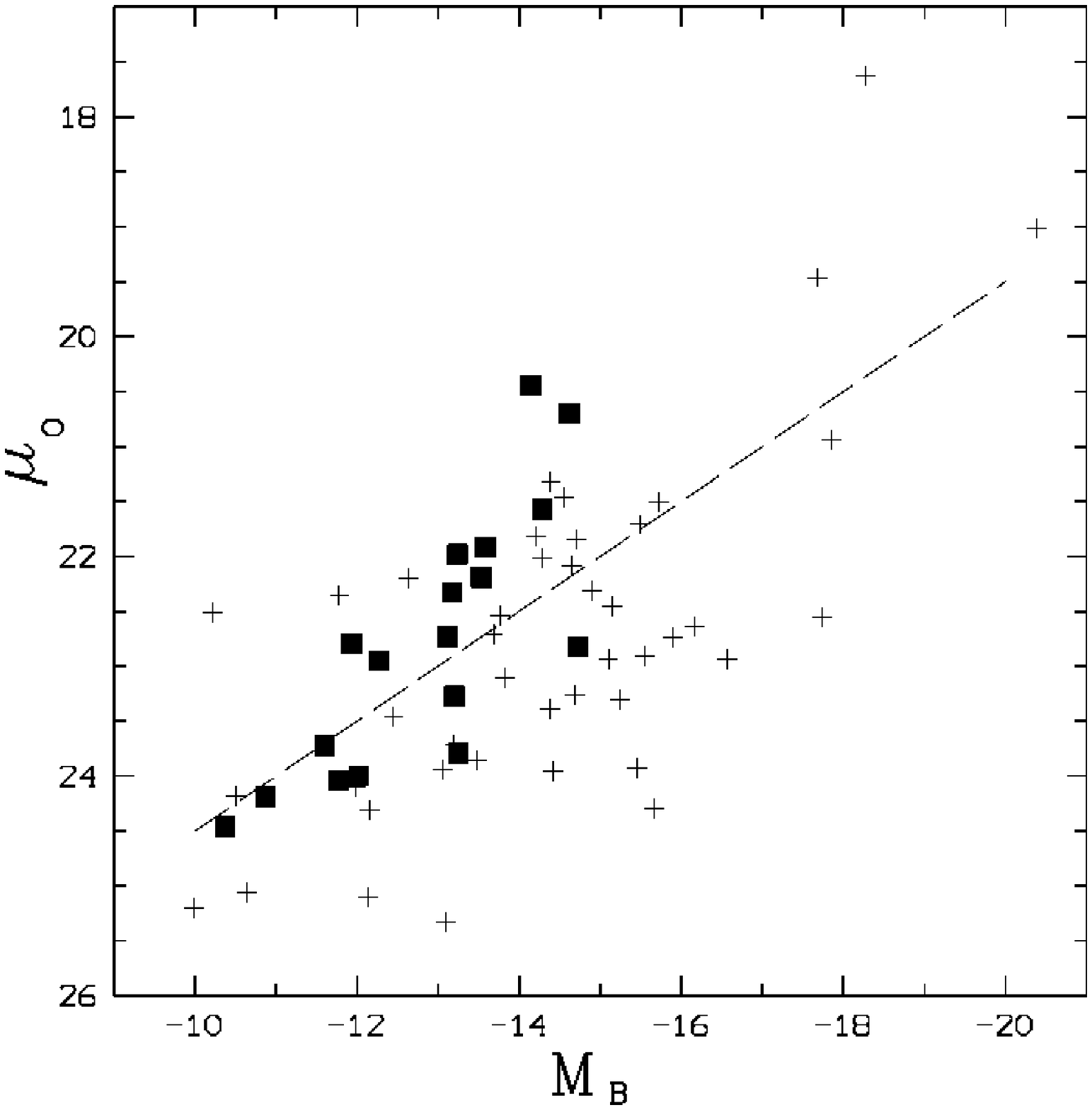,width=8cm}
\psfig{figure=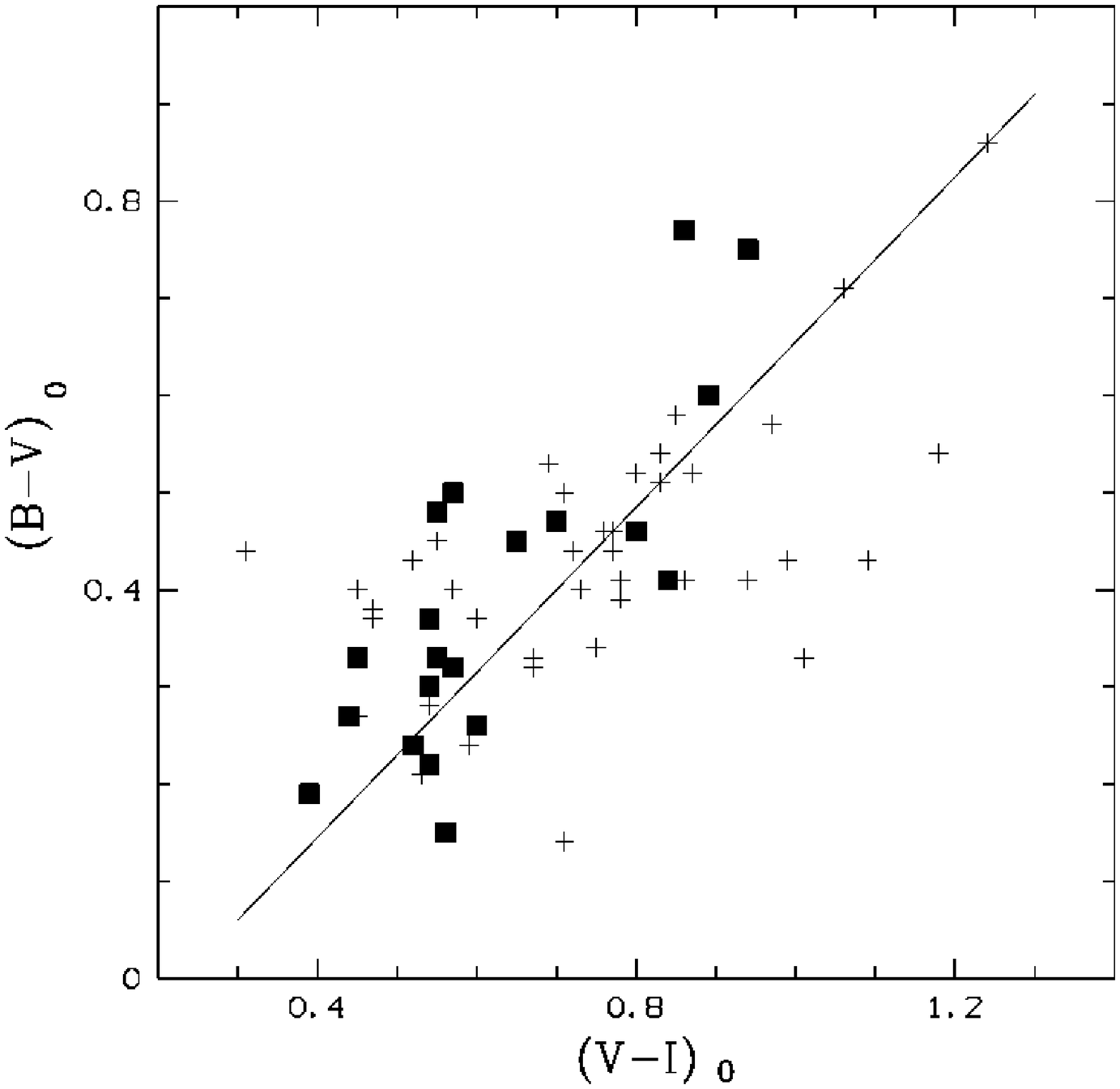,width=8cm}
\caption{The distribution of central surface brightness in $B$ filter
vs. absolute B magnitude (left) and the distribution of the total
colour indexes $(B-V)_0$ vs. $(V-I)_0$. All magnitudes  and colour indexes
are corrected for galactic absorption. The 20 dIrr galaxies from 
the present work are shown with filled squares. The galaxies measured 
in our other studies are shown with crosses.}
\end{figure*}

\section{Notes on individual galaxies}

{\em KKH 46}. This irregular knotty galaxy is one of the most isolated
objects in the LV. The distance is estimated from its radial velocity
with the Hubble constant  H = 72 km/s/Mpc. The apparent magnitude of
KKH 46 g = 18.6 is given in NED from Sloan DSS and it corresponds to a single
blue knot only.

{\em D634-03}. Distance to the object was recently measured via the luminosity
of TRGB (Karachentsev et al. \cite{k06}).

{\em Antlia}. This is a dwarf companion to NGC 3109. Distance measurements
to Antlia via TRGB were made by Aparicio et al. \cite{aparicio} and 
Tully et al. \cite{tully} in fine mutual agreement.

{\em KKH 60}. This is an irregular galaxy of low surface brightness with
an unreliable estimate of radial velocity, +286 km/s (Makarov, Karachentsev,
Burenkov \cite{makarov}).
Judging from its new radial velocity, +1670 km/s, measured at Arecibo
(M. Haynes, personal communication), KKH 60 resides far beyond the LV.

{\em KKs 44 = ESO 320-014}. This dIrr galaxy is situated at the remote
periphery of the CenA/M83 complex. The distance to the galaxy was recently 
determined from the TRGB by Karachentsev et al. \cite{k07}.

{\em KK 144, KK 149, KK 151, and KK 160}. Judging from their radial
velocities, these four dIrr galaxies belong to the CVnI cloud. Three
of them, apart from KK 160, are detected in the H$\alpha$ emission line
(Kaisin \& Karachentsev \cite{kaisinetal}) testifying to active star 
formation in these dwarf systems.

{\em ESO 381-018 and ESO 381-020}. These are two dIrr galaxies on the
outskirts of the CenA/M83 complex. Their distances were recently measured
via TRGB (Karachentsev et al. \cite{k07}).

{\em KKSG 37}. This is a dwarf system of very low surface brightness and
rather uncertain type (dSph or dIrr). The reported HI emission with radial
velocity of +85$\pm$1 km/s (Huchtmeier, Karachentsev, 
Karachentseva \cite{huh}) has apparently 
a local Galactic origin. Judging from the measured colours $B-V$ = 0.82, 
$V-I$ = 1.01, KKSG 37 is a distant dSph system probably located 
in the Virgo Southern extension.

{\em ESO 443--09 = KK 170}. This dIrr galaxy situated at the
outskirts of M83 group has a distance measured from TRGB 
(Karachentsev et al. \cite{k07}).

{\em DDO 155 = GR 8 = VIII Zw 222}. A very nearby dIrr system seen in the
stage of active star formation.

{\em PGC 45628 and PGC 170257}. Two dIrr galaxies on the periphery of
the CenA/M83 complex with distances estimated from their radial velocities.

{\em UGC 8651, UGC 8760, and UGC 8833}. Three dIrr galaxies with accurate
TRGB distances 3.01, 3.18, and 3.19, respectively, in a
group of dwarfs on the near side of the CVnI cloud 
(Tully et al. \cite{tully}).
The R-band profiles of UGC 8651 and UGC 8760 were measured by Swaters
\cite{swat}.

\begin{figure}
\psfig{figure=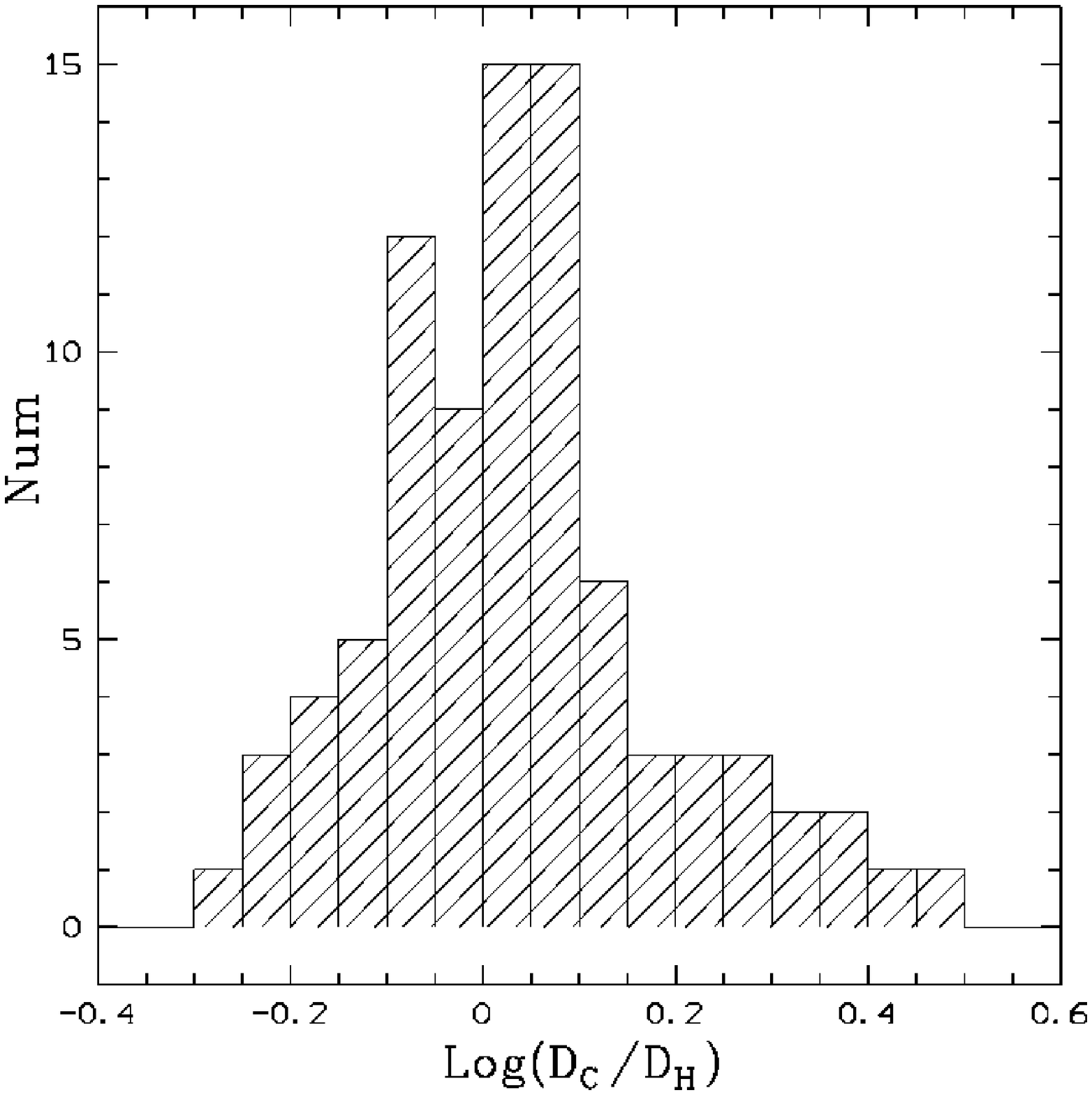,width=8cm}
\caption{The distribution of nearby galaxy diameters from
the Catalogue of Neighbouring Galaxies
relatively to Holmberg diameters for dwarf galaxies studied by us.}
\end{figure}

\section{Discussion and conclusions}
In this work we present imaging and general photometric parameters
of 20 nearby dwarf galaxies.
It is seen in Table~2 and Table~3 that the blue central
surface brightnesses of the 20 irregular galaxies under study lies within
the range of [20.8 - 24.8] mag sq. sec$^{-1}$ and their total
absolute magnitudes occupy the interval of [--10.3, --14.8] mag.
Total parameters of these galaxies follow the common relations for 
dwarf galaxies. The distribution of central surface brightness
and absolute magnitudes for the objects of our sample is presented
in the left panel of Fig.2. In general, the galaxies follow the relation 
$\mu_0(B) \propto 0.5 M_B$ (dashed line) similar to the dwarf
population of the Local Group (Grebel, Gallagher, Harbeck \cite{grebel}). 
Total colour indexes $B-V$
and $V-I$ of the galaxies under study (Fig.2, right panel) follow the relation
$B-V = 0.85 (V-I) - 0.2$ similar to dIrr and dSph galaxies from
our previous paper (Makarova \cite{m98}) with the scatter $\sim$ 0.15$^m$.

The surface brightness profiles of the dwarf Irr galaxies from our sample
are well-fitted by an exponential law in its outer parts, whereas in the
central parts they show light depression. This tendency
is very common for nearby dwarf galaxies, including dIrrs and dSphs
(see, for example, the big series of nearby dwarf galaxies study by 
Bremnes et al. \cite{bremnes98}; Bremnes, Binggeli, Prugniel \cite{bremnes00};
Barazza, Binggeli, Prugniel \cite{barazza}; Parodi, Barazza, 
Binggeli \cite{parodi}). 
In a number of studies of dwarf (mostly spheroidal) galaxies the central 
light depression of surface brightness profiles was fitted
by the Sersic law $I(r) = I_0*exp[-(r/r_0)^n]$ (Jerjen, Binggeli,
Freeman \cite{jerjen},
Graham \& Guzman \cite{graham}). With the Sersic index $n<1$, surface 
brightnesses
flatten toward a constant surface brightness core at the centre and
steepen down from the exponential fall-off at large radii.
One might imagine that the dwarfs would typically be tidally truncated 
and have a cutoff which could have a form described by Sersic $n<1$.
However, most of the profiles considered by us and in the studies 
mentioned above are consistent with an
exponential decay with radius with no hint of a cutoff.
The Sersic parameterization does {\bf not} adequately describe this
characteristic of a flat core (requiring Sersic index $n<1$) but an 
exponential decay at large radii as far as can be followed (described by
Sersic index $n=1$).
Perhaps dwarfs that are in close proximity to a big galaxy
and have had a close encounter might be tidally truncated and
consequently be reasonably described by a Sersic index $n<1$.
However our dwarfs that are for the most part gas-rich and relatively
isolated might have their unmodified `natural'
shape and that shape, evidently, is described by Sersic $n=1$ in their
outer parts.

The surface brightness profiles allow us to determine the
characteristic isophotes of major diameters of
galaxies in the CNG. The distribution of CNG diameters
relatively to Holmberg diameters for 85 dwarf galaxies studied by us
is presented in Fig.3. The diameters measured in the CNG are not homogeneous.
For most bright galaxies (with NGC, UGC, ESO numbers) the diameters
nearly correspond to the level of $\mu_B=25$ mag sq sec$^{-1}$.
Diameters of the low surface brightness objects from the
KK, KKH, KKR, KKSG and KKs lists were measured at fainter
isophotes. Following the data of Fig.3 we conclude that the typical
diameter of a CNG galaxy is close to the Holmberg diameter, exceeding it
by about 10\%. The median isophote of CNG diameters corresponds to
about 26.8 mag sq.sec$^{-1}$ in $B$ according to our data.

{\it Acknowledgments.
This work was partially supported by RFFI grants 07--02--00005
and 08--02--00627 and
grant DFG-RGBR 06--02-04017.}

{}


\begin{thebibliography}{}
\bibitem[1997]{aparicio} Aparicio A., Dalcanton J. J., Gallart C., 
Martinez-Delgado D., 1997, AJ, 114, 1447
\bibitem[2001]{barazza} Barazza F., Binggeli B., Prugniel P., 2001,
A\&A 373, 12
\bibitem[2006]{belokurov} Belokurov V., Zucker D.B., Evans N.W. et al.,
2006, ApJ, 647, L111
\bibitem[1998]{bremnes98} Bremnes T., Binggeli B., Prugniel P., 1998,
A\&AS 129, 313
\bibitem[1999]{bremnes99} Bremnes T., Binggeli B., Prugniel P., 1999,
A\&AS 137, 337
\bibitem[2000]{bremnes00} Bremnes T., Binggeli B., Prugniel P., 2000,
A\&AS 141, 211
\bibitem[1959]{vau} De Vaucouleurs G., 1959, Handbuch der Physik, 53, 275,
Flugge  S.(ed.), Springer, Berlin
\bibitem[2005]{gildepaz} Gil de Paz A., Madore B., 2005, ApJS, 156, 345
\bibitem[2003]{graham} Graham A., Guzman R., 2003, AJ, 125, 2936
\bibitem[2003]{grebel} Grebel E. K., Gallagher J. III, Harbeck D., 2003,
AJ, 125, 1926
\bibitem[1995]{hs95} Hopp U., Schulte-Ladbeck R., 1995, A\&AS, 111, 527
\bibitem[2003]{huh} Huchtmeier W.K., Karachentsev I.D., Karachentseva V.E., 
2003, A\&A, 401, 483
\bibitem[2000]{jarrett} Jarrett T.H., Chester T., Cutri R., et al., 
2000, AJ, 119, 2498
\bibitem[2000]{jerjen} Jerjen H., Binggeli B., Freeman K.C., 2000,
AJ, 119, 593
\bibitem[2008]{kaisinetal} Kaisin S.S., Karachentsev I.D., 2008, 
A\&A, 479, 603
\bibitem[1996]{karachentseva96} Karachentseva V., Prugniel P., Vennik J.,
Richter G., Thuan T., Martin J., 1996, A\&AS, 117, 343
\bibitem[2004]{k04} Karachentsev I., Karachentseva V., Huchtmeier W.,
Makarov D., 2004, AJ, 127, 2031
\bibitem[2006]{k06} Karachentsev I.D., Dolphin A., Tully R.B., et al., 
2006, AJ, 131, 1361
\bibitem[2007]{k07} Karachentsev I.D., Tully R.B., Dolphin A., et al., 2007, 
AJ, 133, 504
\bibitem[2008]{koposov} Koposov S., Belokurov V., Evans N. W., et al., 2008,
ApJ, 686, 279
\bibitem[1992]{landolt} Landolt A., 1992, AJ, 104, 340
\bibitem[2003]{makarov} Makarov D.I., Karachentsev I.D., Burenkov A.N., 2003, 
A\&A, 405, 951
\bibitem[1999]{m98} Makarova L., 1999, A\&AS, 139, 491
\bibitem[1998]{metal98} Makarova L., Karachentsev I., Takalo L.,
Heinamaki P., Valtonen M., 1998, A\&AS, 128, 459
\bibitem[2002]{m02} Makarova L., Karachentsev I., Grebel E., Barsunova O.,
2002, A\&A, 384, 72
\bibitem[2005]{m05} Makarova L., Karachentsev I., Grebel E., Harbeck D.,
Korotkova G., Geisler D., 2005, A\&A, 433, 751
\bibitem[1996]{papaderos} Papaderos P., Loose H.-H., Thuan T., 
Fricke K., 1996, A\&AS, 120, 207
\bibitem[2002]{parodi} Parodi B. R., Barazza F. D., Binggeli B., 2002,
A\&A, 388, 29
\bibitem[1998]{schlegel} Schlegel D.J., Finkbeiner D.P., Davis M.,
1998, ApJ, 500, 525
\bibitem[1999]{swat} Swaters R., 1999, Dark Matter in Late-type 
Dwarf Galaxies, Thesis, Groningen
\bibitem[2006]{tully} Tully R.B., Rizzi L., Dolphin A.E. et al., 2006, 
AJ, 132, 729
\end{thebibliography}
\end{document}